\documentclass[11pt]{article} 
\usepackage{hyperref} \pdfoutput=1
\begin{document}
\title{Mixing Within Patterned Vortex Core}
\author{Hamid Ait Abderrahmane, Mohamed Fayed, Amr Mandour,\\ Lyes Kadem, Georgios H. Vatistas and Hoi Dick Ng
\\ \\\vspace{6pt} Department of Mechanical and Industrial Engineering, \\ Concordia University, Montreal, Canada H3G 1M8, Canada}
\maketitle
\begin{abstract} The video shows the flow dynamics within inner and outer regions of a vortex-core. The observed phenomena mimic a transport process occurring within the Antarctic vortex. The video shows two distinct regions: a strongly mixed core and broad ring of weakly mixed region extending out the vortex core boundaries. The two regions are separated by a thin layer that isolates the weakly and strongly mixed regions; this thin layer behaves as barrier to the mixing of the two regions.  The video shows that the barriers deplete when a swirl of the vortex core increases and the vortex core espouses a triangular pattern.          \end{abstract}
\section{Introduction}
Confined rotating shallow layers of fluid in circular tanks exhibit coherent structures which resemble to a large extent the ones observed in several geophysical and astrophysical flows. These flows involve a solid body rotation of the vortex core and shear layer flow at the outer region. A fluid dynamics video, \href{http://ecommons.library.cornell.edu/bitstream/1813/14091/3/concordia_mpeg2.mpg}{high resolution} or \href{http://ecommons.library.cornell.edu/bitstream/1813/14091/2/concordia_mpeg1.mpg}{low resolution}, is produced in this work to illustrate the flow dynamics within inner and outer regions of such vortex-core. 

The experiments were conducted in a cylindrical Plexiglass container. The internal diameter of the cylinder is 284 mm. A circular disk (252 mm diameter) is placed 20 mm above the bottom of the cylinder. The disk is connected through a vertical shaft to an electric motor. A flywheel is attached to the vertical shaft in order to increase the constancy of the disk rotation. The disk is rotated in the counterclockwise direction, and its speed is regulated by an electronic controller. 

For the flow visualization we used pH indicator thymol blue dye; the color of the indicator is obtained by adding few drops of alkaline and acid solutions. The experiments were conducted with 20 mm of water above the rotating disk; the disk speed varied from 150 to 220 rpm. The camera was placed above the tank that acquires at a rate of 60 frames per second; shutter speed of the camera was set at 1/60 sec. 

Spinning a shallow water layer with a rotating disk in circular tank a vortex was obtained. Depending on the value of the control parameter, disk speed, the vortex core espouses elliptical and triangular shapes. The video shows three regions: a strong mixed vortex core, broad shear layer flow ring and thin layer that separates the two former regions.  

The first part of the video shows that the thin layer, inserted between the vortex core and the outer shear layer, works as barriers and slows down mixing of the two distinct regions. The inner vortex core is the location of strong mixing and expands with time leading at the end to the complete depletions of the thin barrier. The second part of the video corresponds to an increase of the disk speed, the vortex core adopted a triangular pattern. The mixing within the vortex core becomes intense and accompanied with the appearance of a secondary flows or satellites vortices at the apexes of the triangular pattern. The depletion of the thin barrier was very fast and the inner and outer regions mixed. 

\vspace{10pt}
*This work is supported by Natural Sciences and Engineering Research Council of Canada (NSERC).

\end{document}